# Can AI be Consentful?[1]

Giada Pistilli, Bruna Trevelin
Hugging Face

## Abstract

The evolution of generative AI systems exposes the challenges of traditional legal and ethical frameworks built around consent. This chapter examines how the conventional notion of consent, while fundamental to data protection and privacy rights, proves insufficient in addressing the implications of AI-generated content derived from personal data. Through legal and ethical analysis, we show that while individuals can consent to the initial use of their data for AI training, they cannot meaningfully consent to the numerous potential outputs their data might enable or the extent to which the output is used or distributed. We identify three fundamental challenges: the scope problem, the temporality problem, and the autonomy trap, which collectively create what we term a "consent gap" in AI systems and their surrounding ecosystem. We argue that current legal frameworks inadequately address these emerging challenges, particularly regarding individual autonomy, identity rights, and social responsibility, especially in cases where AI-generated content creates new forms of personal representation beyond the scope of the original consent. By examining how these consent limitations intersect with broader principles of responsible AI – including fairness, transparency, accountability, and autonomy – we demonstrate the need to evolve ethical and legal approaches to consent.

## What is consent? Why does it matter in AI?

The concept of consent, fundamental to our understanding of personal rights and autonomy, takes on new complexity in the age of artificial intelligence (AI). As AI systems increasingly influence decision-making in domains such as healthcare, education, and public administration, questions of individual control and informed authorization have become more urgent. These challenges are further intensified in the context of generative AI (a subset of AI capable of producing synthetic content such as text, audio, and images) where data use and output creation significantly complicate traditional notions of consent. While this principle has traditionally governed interpersonal conduct, its application to AI systems exposes significant limitations, particularly considering the vast potential of AI-generated content derived from personal data.

---

[1] This paper an upcoming book chapter which will appear in a collective book to be published by Cambridge University Press, and edited by Marios Constantinides and Daniele Quercia (Nokia Bell Labs).

## Evolution of informed consent

The evolution of consent in the ethics tradition represented a dramatic shift in its first applied context: healthcare ethics. Historical medical practices, exemplified by the ancient Hippocratic Oath[2] from the fifth century BCE, advised physicians to "conceal most things from the patient[3]" (Hasday, 2001) and issue directives with "cheerfulness and serenity" (Hasday, 2001) while withholding information about the patient's condition. This paternalistic approach would eventually give way to a radical transformation in how the doctor-patient relationship is conceptualized and executed. The concept of informed consent emerged as a cornerstone of modern bioethical frameworks, fundamentally reshaping both clinical practice and medical research. Its ascendance during the 20th century marked a shift in healthcare delivery, displacing the traditional physician-centered approach where medical professionals unilaterally determined treatment paths based on their presumed superior judgment.

This evolution toward patient autonomy gained particular momentum in the aftermath of World War II, with the establishment of the Nuremberg Code (1949) serving as a pivotal response to wartime medical atrocities. Since the 1970s, the bioethics literature has extensively explored the prerequisites for *valid* consent, emphasizing three key elements: patient capacity, adequate information, and voluntary choice. These standards, particularly in medical research contexts, have grown increasingly rigorous, making patient self-determination an essential right in both experimental and therapeutic contexts.

Building on these evolving standards, modern medical ethics has developed a precise setting for what constitutes valid informed consent. At its core, informed consent encompasses three essential elements: it must be informed, voluntary, and provided by an individual with adequate decision-making capacity. This framework requires full disclosure of treatment or research protocols, verification of patient understanding, and confirmation that the agreement is made without coercion (Miller, 2009). The legitimacy requirement of informed consent establishes it as a prerequisite for medical interventions – without adequately informed consent, even potentially beneficial procedures that significantly impact a patient's body or privacy become ethically impermissible.

## The transformative moral power of consent

In the ethics tradition, consent operates through two primary mechanisms. First, it acts as a permissive gateway, allowing others access to what would otherwise be protected – in the AI context, this might be personal data, voice recordings, or images used for training. Second, it can function as a binding commitment, creating mutual obligations between parties. However, this dual nature reveals a fundamental challenge in AI systems: while individuals can consent to the initial use of their data, the capabilities of AI create outputs and implications that extend far beyond the scope of the original permission.

---

[2] https://www.pbs.org/wgbh/nova/doctors/oath_classical.html
[3] https://www.nytimes.com/1979/05/01/archives/the-doctors-world-the-doctors-world.html

The normative force of consent is inseparable from its nature. It is not merely a descriptive act (like clicking "I agree" on a data collection notice) but carries inherent moral weight (Miller, 2009). This phenomenon becomes particularly problematic in the context of generative AI, where the act of consenting to data use may enable countless unforeseen applications and representations of personal information. While these traditional frameworks for consent have served as ethical and legal guardrails in medicine, research, and data protection, the emergence of generative AI systems introduces new challenges that strain these frameworks. The fundamental assumptions underlying traditional consent (i.e., specificity, foreseeability, and reversibility) face systematic disruption when applied to technologies capable of creating endless synthetic derivatives from personal data. This tension forms the central concern of our analysis.

## Evolution of legal protection of privacy and consent

Legal approaches to the protection of privacy - and its relation to consent - have evolved along distinct paths in different jurisdictions, reflecting varied historical contexts and philosophical foundations.

In the United States, for example, privacy rights emerged largely from Warren and Brandeis's influential articulation of "the right to be let alone," (Warren, 1890) a concept that gained urgency with the advent of photography and mass-circulation newspapers. This foundation led to a patchwork of sectoral privacy laws built primarily on principles from tort and contract law. Under U.S. tort law, the recognition of privacy violations established that "the right of privacy ceases upon the publication of the facts by the individual, or with his consent," (Warren, 1890) positioning consent as a waiver of protection rather than an affirmative right. The Fair Information Practice Principles (FIPPs), which emerged in a 1973 U.S. government report, helped build a framework that emphasizes transparency, the right to be informed, restrictions on data reuse without consent, the ability to correct records, and accountability for data holders (Solove, 2013). Simultaneously, contract law facilitated the management of personal information exchanges through commercial relationships, with privacy notices functioning as hybrid confidentiality agreements and traditional promise-based contracts (Tschider, 2021). Today, the general U.S. approach to privacy regulation, called "privacy self-management" by Daniel Solove (Solove, 2013), grants individuals rights such as notice, access, and consent to control their personal data.

The European approach, by contrast, mainly developed from the traumatic lessons of World War II (Tschider, 2021), where governments systematically exploited personal information to target citizens based on ethnic, religious, and political status. This experience shaped a rights-based framework codified in instruments from the United Nations Declaration of Human Rights (1948) to the General Data Protection Regulation (GDPR - Regulation 2016/679). European data protection law since at least 1995 has not positioned consent as the sole legal basis for data processing, instead establishing multiple legal grounds that attempt to balance individual rights with organizational interests. Nevertheless, consent remains a central pillar in the European framework. According to the GDPR, today one of the most protective frameworks of data protection in force, consent to the processing of one's personal data should be given by "a clear affirmative act establishing a freely given, specific, informed and unambiguous indication of the data subject's[4] agreement" (Recital 32). It should cover all processing activities carried out for the same purpose(s) and, when the processing has multiple purposes, consent should be given to each and all of them. It can also be withdrawn at any given time.

# Why is consent challenging in AI?

In conventional contexts, consent is bounded by clear parameters: it is given for a particular purpose, with a reasonable understanding of potential outcomes, and with the possibility to be withdrawn. However, the application of this setting to AI systems reveals fundamental incompatibilities.

Traditional consent frameworks have evolved in contexts where boundaries remain relatively stable and outcomes reasonably predictable. In medical settings, a patient consents to a specific procedure with known risks and benefits, performed by identifiable practitioners, within a defined timeframe. In research participation, subjects authorize specific experimental protocols with established methodologies and clearly articulated purposes. In data protection regimes like the GDPR, individuals theoretically consent to particular processing activities for enumerated purposes by identifiable data controllers. These conventional models share common structural elements: specificity of purpose, clarity of scope, identifiable responsible parties, temporal boundaries, and meaningful withdrawal mechanisms (GDPR - Regulation 2016/679). Each element functions as a constraining parameter that makes the consent meaningful as an exercise of autonomous choice.

When applied to AI systems, however, these structural elements encounter unique challenges that call into question whether "consent" in its traditional sense remains conceptually coherent or actually applicable. AI systems are spread across many actors and stages, which makes it hard to keep consent meaningful. These incompatibilities are not only practical and technical implementation difficulties, but represent conceptual misalignments between traditional consent frameworks and the technological realities of contemporary AI systems, particularly those with generative capabilities.

Therefore, in the AI context, consent faces three distinct challenges that strain traditional definitions.

## The scope problem: informed consent becomes impossible

First, the *scope problem*: consent can only be applied when people are aware that their data is being used. In AI, vast amounts of data are being scraped and repurposed without individuals' knowledge, notably publicly available information. Scraping[5] essentially ignores privacy considerations, giving no notice to people whose data are being used, much less so the possibility of giving or withdrawing consent for this use.

---

[4] Data subject is an identified or identifiable natural person to whom personal data may be related. Throughout the article we may use some of terms defined in the GDPR, such as "data subject", "personal data", "controller".

But in a scenario where there is awareness of such use, while individuals may consent to the initial collection and use of their data, the generative capabilities of AI systems can produce countless derivative works and representations that extend far beyond reasonable foreseeability. For instance, a person consenting to voice sampling for a specific application cannot anticipate or evaluate all possible synthetic utterances their voice might produce in perpetuity.

To give another example, consider the case of Clearview AI[6],[7],[8]. Clearview AI became a major player in facial recognition technology by collecting billions of images from social media, online profiles, and photography websites. The company gathered these photos without informing any individuals, ignoring all terms of service and potentially applicable law, and notably without obtaining consent. They went on to use images that people never expected to be included in a large-scale AI system as a facial recognition tool that quickly became widely used for law enforcement and government surveillance.

This unauthorized access to and use of data is not unique to Clearview AI but represents a broader pattern in AI development. Desai and Riedl (2024) highlight how AI companies often gather copyrighted material for training without permission, creating tension between technological advancement and respecting ownership and consent rights. Even though consent in data protection and consent as authorization in copyright may differ, this pattern of behavior demonstrates how AI systems systematically undermine the informed aspect of informed consent through both data-gathering practices and their potential outputs.

Therefore, this scope problem fundamentally undermines the "informed" aspect of informed consent. Consider a voice actor[9] who provides recordings for a text-to-speech system with explicit permission for use in audiobook narration. The same underlying model, once trained, could potentially generate speech patterns imitating that actor delivering hate speech, making political endorsements, or providing financial advice – all contexts entirely separate from the original consent domain. The endless combinations enabled by generative systems effectively render comprehensive foreseeability impossible, as the potential outputs grow exponentially with each new application and context. In trying to deal with this reality, the Screen Actors Guild - American Federation of Television and Radio Artist (SAG-AFTRA) has struck AI voice agreements with Replica Studios[10] and Narrativ[11], following near-unanimous support for a strike authorization over AI job threats and safety concerns. According to the deals[12], actors can opt out or in and have some say over how their voices are used, but it remains unclear to what extent this can protect them.

---

[5] Web scraping is the process of automatically mining data or collecting information from the World Wide Web. This automatic mining can respect a protocol called "robots.txt", a simple text file included in websites to ask that the content not be scraped. However, this protocol is not legally binding, so respect to it remains largely optional. (Solove and Hartzog, 2024, p. 8)
[6] https://www.nytimes.com/2020/01/18/technology/clearview-privacy-facial-recognition.html
[7] https://gpdp.it/web/guest/home/docweb/-/docweb-display/docweb/9751323
[8] https://www.buzzfeednews.com/article/ryanmac/senators-markey-wyden-clearview-ai-facial-recognition
[9] See what happened during the Hollywood strike: https://www.nytimes.com/live/2023/07/13/business/actors-strike-sag

Moreover, the scope problem extends beyond individual instances to enclose systemic effects that emerge when personal data becomes part of larger training corpora. When a person's data is incorporated into a generative system, it practically contributes to the model's overall capability to represent and generate content resembling that individual. This phenomenon creates what might be termed "representational drift", where the connection between original consent and subsequent representations becomes increasingly tenuous as the model evolves through additional training or fine-tuning processes.

The legal frameworks governing traditional consent rely heavily on the principle of specificity – that individuals understand and agree to particular uses of their information or likeness. Yet generative AI systematically violates this principle through its inherent flexibility. Unlike traditional media where content remains fixed and bounded, generative outputs remain perpetually malleable, capable of being reformulated into new contexts that the original data subject could not have reasonably anticipated or evaluated during the initial consent process.

On a more positive note, some organizations have attempted to address the scope problem through more granular consent mechanisms. For instance, the Partnership on AI's ABOUT ML (Annotation and Benchmarking on Understanding and Transparency of Machine Learning Lifecycles)[13] initiative promotes documentation standards that require detailed tracking of data provenance and usage limitations. Companies like Apple have implemented differential privacy techniques[14] that allow statistical learning while maintaining individual data protection. These approaches represent steps toward more ethical and lawful AI development, though they still struggle with the fundamental limitations of consent in generative contexts.

Within the scope problem, we can think about *data generation*. Traditional consent and privacy protections operate and typically inform individuals about the data *collected* about them but often fail to address data that is generated about them. The potential for privacy violations increases when inferences drawn from data collected reveal sensitive details people may not want to disclose. A well-known example of this occurred with Target[15], an American mass-market retail company, which used an algorithm to identify pregnant women based on their shopping habits – the algorithm aimed to recognize pregnancy early and promote baby-related products. In one case, a father complained to Target after his teenage daughter began receiving baby product ads, assuming they were sent by mistake or that Target was incentivizing pregnancy[16]. Target's algorithm identified pregnancy by recognizing specific purchases, such as unscented products, vitamins, and cotton balls, which were common among pregnant shoppers, and made ads accordingly. The father later discovered that his daughter was indeed pregnant. The issue here is that the algorithm could infer sensitive health information from seemingly less sensitive data, like shopping patterns, and the factors used to make such inferences are not easily anticipated. People cannot understand or foresee what types of information can be inferred using the data they disclose.[17]

---

[10] https://www.sagaftra.org/sag-aftra-and-narrativ-announce-new-agreement
[11] https://www.sagaftra.org/sag-aftra-and-replica-studios-introduce-groundbreaking-ai-voice-agreement-ces
[12] https://www.forbes.com/sites/virginieberger/2024/08/21/sag-aftras-ai-deal-a-5-billion-gamble-on-the-future-of-voice-acting/&sa=D&source=docs&ust=1742816906097772&usg=AOvVaw1Kn2bMGWkONPNgernYM63t
[13] https://partnershiponai.org/wp-content/uploads/2021/08/ABOUT-ML-v0-Draft-Final.pdf
[14] https://www.apple.com/privacy/docs/Differential_Privacy_Overview.pdf
[15] https://www.nytimes.com/2012/02/19/magazine/shopping-habits.html

## The temporality problem: consent cannot be meaningfully withdrawn

Second, the *temporality problem*: traditional consent operates within a discrete timeframe with clear boundaries, whereas AI systems create an open-ended relationship between data subjects and their representations. Unlike a medical procedure with a defined beginning and end, AI-generated content derived from personal data can persist, evolve, and multiply indefinitely, rendering the traditional notion of revocable consent problematic at best and meaningless at worst.

The temporality problem manifests in several dimensions that collectively challenge conventional consent paradigms.

First, there is the issue of persistence: once personal data enters a training dataset, extracting its influence from the resulting model becomes technically hard, if not impossible at a large scale. Even if a data subject were to formally withdraw consent after model training, their "digital trace" remains embedded within the model's parameters, continuing to influence outputs without clear pathways for removal. Therefore, there is a fundamental asymmetry here: consent can be given at a discrete moment, but cannot be meaningfully withdrawn in the same discrete fashion without complete model retraining. Even with retraining, there may be residual effects from previously learned patterns, and the practical barriers of cost, time, and technical complexity make immediate and complete removal of influence nearly impossible. That means that the right to withdraw consent, or the right to ask for deletion of one's personal data, as available according to the GDPR, are very likely not respected within the AI lifecycle[18].

---

[16] https://www.forbes.com/sites/kashmirhill/2012/02/16/how-target-figured-out-a-teen-girl-was-pregnant-before-her-father-did/

[17] The growing volume and connection of various types of data make it increasingly possible to predict people's health and behavior, blurring the line between health and non-health data. Since even non-health data can reveal sensitive health information, Shnebel, Elger and Shaw argue that the definition of "health data" should be expanded to include inferred, indirect, and future health-related data. (Schneble CO, Elger BS, Shaw DM. 2020)

[18] The French data protection authority - CNIL (Commission nationale de l'informatique et des libertés) published in 2024 "AI how-to sheets" with recommendations for the development of AI systems and the creation of datasets used for their training, when they involve personal data. In one of these sheets (Respect and facilitate the exercise of data subjects' rights), they recommend that model providers

Second, there is the evolutionary dimension of AI systems, whereby models trained on personal data continue to develop through additional training iterations, fine-tuning processes, and architectural modifications. Each evolution potentially transforms how the original data influences outputs, creating temporal distance between the initial consent act and subsequent representations. This progressive divergence means that even if consent were adequately informed at the outset (itself a problematic assumption given the scope problem), its validity diminishes over time as the system evolves beyond its original parameters. Also, as AI models evolve, companies deploying them may also struggle to obtain consent from data subjects. This is because they often lack access to the original datasets used for training, making it difficult to track, verify, or re-seek consent from individuals whose data contributed to the AI's development – particularly as data provenance becomes obscured through common preprocessing practices (Muller & Strohmayer, 2022; Muller, 2024).

Third, the multiplication effect compounds both persistence and evolution issues. When AI-generated content derived from personal data proliferates across digital platforms, each instance potentially serves as new training data for future models, creating repetitive use and re-use of data over time that extend indefinitely beyond the temporal boundaries of initial consent. This creates what might be termed "consent decay": the progressive weakening of the connection between original authorization and subsequent uses as representations multiply across time and technological contexts (Custers, 2016).

Traditional legal frameworks for consent presuppose temporal boundedness: actions are authorized for specific durations, after which authorization expires or requires renewal. Medical consent, for instance, doesn't imply perpetual authorization for recurring procedures, nor does consent to participate in a research study extend indefinitely to future studies. Yet, AI systems fundamentally violate this temporal boundedness through their evolving nature. This temporal disconnect raises serious questions about intergenerational ethics and justice. When consent given today enables AI-generated representations that may persist for decades or even centuries, future impacts on reputation, privacy, and dignity become impossible to evaluate meaningfully at the time of initial consent.

To address aspects of the temporality problem, some researchers have developed technical tools like "machine unlearning" (Bourtoule, 2021) methods, which aim to remove the influence of specific data from trained models. For example, the SISA (Sharded, Isolated, Sliced, and Aggregated) training framework divides data into manageable shards that allow for more selective removal of data. While promising, these approaches face some computational challenges at scale and cannot fully resolve the persistence of derived or synthetic content once it has been generated and distributed.

periodically retrain their models if they have their training datasets in order to exercise the rights of subjects, like erasing or correcting their data.This would be done "whenever it is not disproportionate to the rights of the controller" and ideally within the GDPR timeframe for complying with these subject's requests, which is two months maximum. However, retraining models - specially in that frequency - is very resource consuming economically and environmentally, besides partially inefective as discussed in this section. So there remains a question on proportionality and effectivenness of this recommendation, especially when it comes to open source models not backed by commercial activities.   (Available at: https://www.cnil.fr/fr/ai-how-to-sheets).

## The autonomy trap: consent undermines future self-determination

Third, the *autonomy trap*: traditional consent settings presume that giving consent is an exercise of individual autonomy. However, in AI contexts, the act of consent itself often undermines future autonomy in ways that cannot be fully appreciated at the time of consent. Namely, when individuals consent to data use for AI training, they are effectively authorizing systems that may later restrict their options, shape their choices through predictive algorithms, or create synthetic representations that influence how others perceive them. This creates a paradoxical situation where the exercise of autonomy through consent enables its own future limitation in ways that traditional consent frameworks cannot account for.

The autonomy trap operates through several distinct mechanisms that collectively challenge conventional understandings of consent as autonomy-enhancing. First, there is the *representational feedback loop*, whereby consenting to data use enables the creation of algorithmic representations of oneself that subsequently influence how others perceive and interact with the consenting individual. For example, when a person consents to social media data collection, they enable recommendation systems that not only shape what content they see but also how their profile is presented to others, potentially reinforcing certain aspects of their identity while suppressing others. Take the example of the pregnant teen discussed above, how from seemingly non-sensitive data (shopping patterns), an algorithm could predict sensitive health data (pregnancy) and expose it to other people (mail with ads that the woman's family accessed) and inform how the company sees the person. This creates a recursive cycle where consented data use progressively narrows one's self-presentation options in digital spaces.

Second, the predictive constraint effect emerges when AI systems use consented data to make predictions that limit individual options. Consider lending algorithms trained on financial histories: by consenting to data collection, individuals enable systems that later categorize them according to risk profiles, potentially restricting future financial autonomy through differential access to credit, housing, or employment opportunities. The initial consent, though ostensibly autonomy-affirming, enables algorithmic systems that may later constrain autonomy in domains far removed from the original context.

Third, the synthetic representation problem arises as AI systems increasingly generate content that mimics individuals' voices[19], writing styles, or visual likenesses. When a person consents to including their data in training sets, they enable the creation of synthetic versions of themselves that may act, speak, or appear in ways contrary to their values or preferences. Unlike traditional reputation management, where individuals can directly influence how they are perceived, these AI-generated representations exist beyond individual control while still being attributed to the original person. This creates what philosopher Deborah Johnson might term "extended identity" problems, where aspects of one's identity become externalized in algorithmic systems yet continue to affect personal autonomy (Johnson, 2006).

---

[19] E.g., see the Scarlett Johansson vs OpenAI dispute:
https://www.theguardian.com/technology/article/2024/may/20/chatgpt-scarlett-johansson-voice

This autonomy trap fundamentally challenges consent's traditional function as a mechanism for preserving and expressing individual self-determination. In conventional contexts, consent operates as a boundary-setting tool, allowing individuals to control access to their person, property, or information. Yet in AI systems, consent increasingly functions as a something that blurs boundaries, where initial authorization enables progressive erosion of future autonomous choice.

## Does Consentful AI create conflicts with other Responsible AI principles?

The challenges of consent in AI intersect with and amplify other core principles of responsible AI development and deployment. These interconnections create a difficult ethical ecosystem where addressing consent gaps involves engaging with broader normative frameworks governing AI technologies.

### Fairness and distributive justice

The consent gap interacts most prominently with fairness considerations, particularly distributive justice (Rawls, 1971; Nozick, 1973). When AI systems generate outputs based on training data that includes personal information, profound questions arise regarding who bears the risks of unexpected representations versus who reaps the benefits. This tension manifests in several distinct dimensions of fairness.

First, there is *representational fairness*. Consider large language models trained on vast corpora of internet text, which inevitably contain personal narratives, writings, and discussions from countless individuals who never explicitly consented to their inclusion. The representational burden falls disproportionately on populations with significant online presence but limited legal or technical resources to protect their data – often younger users[20], marginalized communities, and those from regions with less robust privacy regulations. For instance, chatbots trained on Reddit conversations may replicate distinctive writing styles and perspectives of active community members without their knowledge, effectively creating "synthetic personas" based on real individuals' expressive patterns.

Second, *economic fairness* concerns arise when examining who captures value from AI systems trained on personal data. When a generative AI company develops a voice synthesis product using recordings from thousands of individuals – perhaps obtained through seemingly unrelated consumer products like virtual assistants or voice-activated services – the resulting economic benefits flow primarily to the company and its customers, not to the original voice contributors whose data made the technology possible. This creates asymmetric value extraction where consent, even when nominally obtained through terms of service, fails to ensure equitable distribution of benefits.

---

[20] Throughout this chapter, we use the term "data subject" when referring to individuals in legal or regulatory contexts (e.g., under the GDPR). In other contexts, we use "individual" or "user" interchangeably for readability, particularly when discussing ethical, technical, or experiential aspects of AI systems.

Third, *power fairness* emerges as particularly problematic. Marginalized communities often face what scholars term "manufactured consent" – formal compliance with legal consent requirements without meaningful choice or negotiating power. For example, facial recognition systems disproportionately trained on images of underrepresented individuals from public datasets create heightened surveillance risks for these communities, who simultaneously have the least institutional power to withhold consent or influence how their biometric data is used. The historical parallel to medical research conducted on vulnerable populations without meaningful consent highlights the persistent danger of technological development reproducing existing power imbalances.[21]

A concrete example illustrates these fairness dimensions: when a healthcare AI system is trained on patient records from public hospitals serving predominantly lower-income communities, the resulting technology may be deployed in private facilities serving wealthier populations. Those bearing the privacy risks (through inclusion in training data) differ systematically from those receiving the benefits (improved healthcare through AI assistance), creating a distributive justice problem that extends beyond individual consent questions to structural fairness considerations. Moreover, some organizations are attempting to address these fairness concerns through compensation frameworks for data contributors. For example, Datavant[22], a health data company, has pioneered programs where patients can opt-in to sharing their medical data and receive compensation when their information contributes to research outcomes. Similarly, platforms like Ocean Protocol[23] create data marketplaces with transparent tracking of data usage and value attribution. However, these approaches have yet to scale effectively for the massive datasets used in training foundation models, where individual contributions become nearly impossible to isolate and compensate.

## Transparency and explainability

Transparency presents another critical intersection with consent challenges, creating what might be termed the "transparency paradox" in AI consent. Meaningful consent requires adequate information about potential outcomes, yet the complexity and opacity of many AI systems make such transparency practically impossible. This paradox arises because the very technical sophistication that enables AI's transformative capabilities simultaneously undermines the possibility of fully informed consent.

---

[21] In her book "Technocolonialism", Mirca Madianou explores the rapid digitalization and datafication of humanitarian operations and how AI and digital inovation are being promoted as solutions to complex social and economic challenges. She talks about how biolmetric technologies are being used in refufee registrations, blockchain technologies for cash transfers and cryptocurrencies for cash disbursement, algorithms for deciding who will be included in aid distribution, AI for predicting refugee flows and future crisis and so on. There remains a question on the possibility of consenting to all of this data gathered from the populations that are affected by these situations. (Madianou, 2025)

[22] https://www.healthra.org/wp-content/uploads/2018/10/Datavant_De-Identifying-and-Linking-Structured-Data-Whitepaper_vF.pdf

[23] https://phemex.com/academy/what-is-ocean-protocol

The transparency challenge operates at multiple levels. At the *technical level*, even developers themselves often cannot fully explain why advanced models like transformer-based systems produce specific outputs, creating what AI researchers call the "black box problem." If those creating the systems cannot predict or explain outputs with certainty, providing transparent information to data subjects becomes fundamentally compromised. For example, a person consenting to include their writing in a training corpus cannot be meaningfully informed about how their stylistic elements might manifest in future AI-generated texts when even system architects cannot predict such emergent properties precisely. In GDPR terms, this lack of predictability raises concerns about whether consent can truly be *specific*, which would require that individuals be informed about the precise purposes and uses of their data. If the downstream impacts of AI training are inherently uncertain, the specificity of consent becomes questionable, potentially undermining its validity.

At the *cognitive level*, transparency faces human comprehension limitations. The technical details of neural network architecture, embedding spaces, and transformer mechanisms exceed the specialized knowledge of most individuals. Even if extended technical documentation were provided, most data subjects would lack the expertise to translate this information into meaningful understanding of potential implications for their privacy, reputation, or autonomy. This creates an inevitable gap between theoretical transparency (information availability) and effective intelligibility (meaningful comprehension). In GDPR terms, this gap raises concerns about whether consent can truly be *informed*, which would require that individuals understand the nature, scope, and consequences of data processing. If the complexity of AI models renders their functions incomprehensible to the average person, then even the most detailed disclosures may fail to achieve meaningful transparency. Without genuine comprehension, data subjects may give consent without fully grasping the potential risks, calling into question the legitimacy of such consent under GDPR standards.

At the *scale level*, the sheer volume of potential outputs from generative systems makes comprehensive disclosure impossible. Consider an image synthesis model trained on personal photographs: the potential combinations of visual elements, styles, and contexts are effectively infinite. No consent process, however detailed, could meaningfully disclose all possible ways an individual's visual characteristics might appear in future generated images. This unpredictability challenges whether consent can truly be *unambiguous* in GDPR terms. If individuals cannot foresee the vast and varied ways their data might be used in AI-generated outputs, their agreement risks being broad and indeterminate.

A practical illustration comes from text-to-image systems trained on artists' works, which do not necessarily involve privacy but show how acceptance and consent to use any type of data work. Despite technical documentation revealing training methodologies, individual artists cannot reasonably predict how their distinctive techniques might manifest across billions of potential image prompts. Some may find their artistic "signature" applied to generated content they find objectionable or that damages their professional reputation. The transparency provided during consent processes – typically through generic statements about "training AI models" – falls dramatically short of enabling meaningful evaluation of such specific risks. Moreover, organizations implementing lawful AI transparency include Canada's Algorithmic Impact Assessment (AIA) framework[24], which requires government agencies to evaluate and document AI system risks before deployment. Private sector examples include IBM's AI FactSheets[25], which provide standardized documentation of AI system characteristics, testing results, and intended uses. When properly implemented, these tools help address the transparency paradox by making AI characteristics more accessible to data subjects, though they cannot fully resolve the inherent unpredictability of generative systems. Recent frameworks such as Bogucka et al.'s (2024) responsible AI reporting structure and Stahl et al.'s (2023) review of AI impact assessments stress the importance of connecting transparency practices with concrete, auditable risk documentation. However, as Xia et al. (2023) note, most current approaches still struggle to operationalize risk in a way that is actionable across real-world deployments.

## Accountability and responsibility

Consent frameworks traditionally establish clear lines of responsibility – the entity receiving consent bears accountability for honoring its boundaries. However, AI systems fundamentally disrupt this clarity through what might be termed "responsibility diffusion." When AI-generated outputs exceed the scope of original consent, determining who bears responsibility for resulting harms[26] becomes legally and ethically complex.

This diffusion operates through several mechanisms. First, *causal complexity* makes attribution difficult. When a synthetic voice trained on thousands of recordings produces harmful content mimicking a particular speaker, establishing a direct causal link between specific data inputs and problematic outputs becomes technically challenging, if not impossible. This technical ambiguity creates legal uncertainty about responsibility assignment. The GDPR's definitions of *controller* and *processor* struggle to account for the layered and iterative nature of AI development. When data passes through multiple stages, collected by one entity, processed by another, and incorporated into evolving models by yet another, it may become difficult to establish who bears the responsibility for obtaining valid consent – should they all obtain consent for each stage? If no single actor has full visibility into how data will be used across successive AI iterations, then assigning accountability for gathering consent becomes increasingly difficult, potentially leaving data subjects without a clear avenue for redress.

---

[24] https://open.canada.ca/data/en/dataset/5423054a-093c-4239-85be-fa0b36ae0b2e
[25] https://dataplatform.cloud.ibm.com/docs/content/wsj/analyze-data/factsheets-model-inventory.html
[26] Assessing harm from privacy violations is inherently difficult because individuals are often asked to weigh long-term, cumulative risks against short-term benefits at the moment their data is collected. Many privacy harms are subtle, dispersed, and only emerge over time through the aggregation of data, making it hard for individuals to foresee their impact. Moreover, the harms aren't just personal; privacy has broader social and cultural importance, supporting intellectual freedom, democracy, and societal well-being. Yet current models of privacy self-management place too much responsibility on individuals and fail to address the collective and societal harms that can result from widespread data practices (Solove, 2013).

Second, *institutional fragmentation* distributes responsibility across organizational boundaries. The entity collecting initial consent may differ from those processing data, training models, deploying systems, or using generated outputs. For instance, a social media platform may obtain consent for data collection, sell this data to third-party AI developers, who then license their models to content creation platforms, creating multiple degrees of separation between the original consent relationship and ultimate usage contexts. This fragmentation creates accountability gaps where no single entity takes responsibility for consent boundaries. Moreover, jurisdictional differences further complicate accountability, as entities involved in data collection, processing, and AI deployment may operate under different legal frameworks, making it unclear which affected people have which rights, which regulatory authority has oversight, and how enforcement applies across borders.

Third, *temporal diffusion* of responsibility occurs as systems evolve. Initial developers establish certain parameters, but subsequent modifications, fine-tuning processes, and deployment decisions potentially transform how personal data influences outputs. This creates discontinuity between those initially committed to respecting consent boundaries and those subsequently influencing system behavior, further complicating accountability structures. Additionally, this temporal diffusion raises challenges for *withdrawing* consent, a right under the GDPR. If personal data is incorporated into AI models that undergo continuous adaptation, it may be practically impossible to extract or erase specific data traces. Once data has influenced model weights or been disseminated across derivative systems, the ability to meaningfully retract consent diminishes, undermining the GDPR's requirement that withdrawal be as easy as giving consent.

A tangible example emerged with voice assistant technologies. Users consented to voice recordings ostensibly for service improvement, only to discover years later that human contractors were reviewing these recordings, including intimate conversations accidentally captured. The fragmentation of responsibility between the technology company, its contractors, and the individuals reviewing the recordings created accountability confusion that undermined the original consent relationship. Similar scenarios multiply exponentially with generative AI systems, where the distance between initial data collection and ultimate usage contexts grows increasingly remote.

## Autonomy and self-determination

The autonomy principle, central to both ethical AI and legitimate consent, faces unique challenges in generative contexts. While autonomy generally entails control over one's self-representation and personal data, generative AI can create synthetic versions of individuals that exist beyond their control yet may significantly impact their reputation, privacy, and dignity.

This autonomy challenge manifests through several mechanisms. First, *representational autonomy* faces unprecedented threats from synthetic media. When AI systems can generate convincing images, voices, or writing styles mimicking real individuals, the traditional boundaries of self-representation dissolve. Unlike conventional media where unauthorized portrayals might constitute defamation or right of publicity violations, generative outputs create a new category of

"synthetic representation" that exists in a legal and ethical gray area. For example, deepfake videos can place an individual's likeness in contexts they would never consent to, yet proving harm under existing legal frameworks remains challenging.

Second, *informational autonomy* becomes compromised when AI systems extract patterns from personal data that individuals themselves may not perceive. Consider healthcare AI analyzing patient records: the system might identify correlations between seemingly unrelated personal characteristics that patients themselves would consider intimate and would not have consented to reveal if asked directly. This creates what philosopher Helen Nissenbaum terms "privacy in context" violations, where information appropriately shared in one context is repurposed in ways that undermine autonomy in another.

Third, *decisional autonomy* faces threats when AI systems make predictions or generate content that influences how others perceive and interact with an individual. For instance, AI-generated profiles synthesizing public information about a person may influence hiring decisions, social opportunities, or public perception without that person's knowledge or consent. This creates a form of algorithmic determination that constrains individual autonomy through synthetic representations beyond one's control.

The case of voice cloning technology offers a compelling illustration. Professional voice actors who consented to recordings for specific projects have discovered their synthetic voices being used for unauthorized commercial applications, political messages, or inappropriate content. Their autonomy becomes compromised not merely through unauthorized use of original recordings (which traditional consent frameworks might address) but through the creation of entirely new synthetic utterances they have never spoken yet, which listeners attribute to them. This represents a novel form of autonomy violation that traditional consent frameworks cannot adequately conceptualize or address.

## What makes it so hard to solve?

There are so many factors to consider when addressing the challenge of consent in AI, making the situation highly complex. As discussed above, fairness and distributive justice issues arise because individuals do not have equal power, representation, or economic standing when interacting with AI systems. Transparency and explainability issues create further barriers: AI models are often too complex to fully understand, and affected people may have comprehension limitations on a technical, cognitive and scale levels, preventing them from making truly informed decisions.

The empirical landscape confirms these theoretical challenges. In an audit of web consent signals, Longpre et al. (2024) documented not only a rapid rise in data use restrictions but also considerable inconsistencies in how these preferences are expressed. Their analysis revealed misalignments between robots.txt specifications and Terms of Service agreements, alongside troubling asymmetries in how different AI developers are restricted, highlighting the technical and organizational obstacles to implementing meaningful consent frameworks.

AI decisions involve multiple stakeholders, leading to causal complexity, institutional fragmentation, and temporal diffusion concerning accountability and responsibility regarding consent in AI. With no single entity fully responsible, it becomes difficult to enforce consent or address violations. Also, autonomy and self-determination are at risk; many people lack control over how their data is collected and used, and AI-driven interactions often affect people's representational, informational, and decisional autonomy.

These challenges make it difficult to create systems where individuals can genuinely give or withhold consent in a meaningful way. There are no easy, ready to employ solutions to address consent in AI.

## How is consent applied in practice?

These developments expose structural weaknesses in consent-based protection. Even though consent is not necessarily the only legal basis for processing of personal data (at least in the EU, the GDPR provides for other bases like legitimate interest, see article 6), it is the most important or most used basis for processing. However, consent places an unrealistic burden on individuals to manage privacy choices across thousands of services and platforms - be it for AI or other purposes. There are increasing discussions on what is commonly called "consent fatigue" or "consent desensitization", which occurs when individuals become overwhelmed and disengaged due to the continuous need to approve data collection and processing requests and may give consent easily, devaluing consent and lowering the level of data protection in the long term (Bergemann, 2018).

Also, the contractual foundations of consent often prove hollow. Privacy notices can change frequently, and platforms often make no binding promises regarding how user data will be used. This renders consent meaningless, as it lacks specific content. Without detailed and consistent agreements, the consent individuals give is more about compliance with terms than genuine control over their data.

Moreover, there is a significant information asymmetry and power imbalance in the relationship between consumers and platforms. Most consumers have little to no real choice or control over the data they provide. Genuinely "freely given" consent is nearly impossible.

One interesting case that shows this involves Facebook. On February 7, 2019, the German competition authority, Bundeskartellamt, concluded an investigation into Facebook's data collection practices, and determined that Facebook held a dominant position in the German social media market and had abused this position by engaging in excessive data collection.

Facebook operates in a "zero-price market," where users do not pay with money but instead provide their personal data, which is used to generate revenue through targeted advertising. Facebook's terms of use require users to accept data collection not only from its own platform but also from other services within the Meta group (such as WhatsApp and Instagram) and third-party websites and apps. The Bundeskartellamt ruled that German users should not be forced to consent to this type of data collection as a condition for using Facebook. It viewed this practice as an abuse of market power, as users would not be subjected to such extensive tracking if Facebook did not dominate the market.

The decision was significant because it treated a violation of the GDPR, enabled by Facebook's market dominance, as an infringement of competition law. It is a different legal domain - competition law - being called to intervene and help enforce privacy protection, unfortunately turned meaningless when users are faced with such information asymmetries and power imbalances[27]. This suggests a critical "consent gap" in current legal frameworks: while individuals can theoretically consent to initial data collection and specific uses, they cannot meaningfully consent to the numerous potential outputs their data might enable through generative AI systems. This gap calls into question whether consent, as currently conceptualized in legal systems, remains an adequate foundation for protecting individual autonomy in the age of generative AI.

These interconnections demonstrate that addressing the consent gap in AI requires a holistic approach that considers how consent functions within a broader ecosystem of ethical principles. Rather than treating consent as a standalone issue to be resolved in isolation, effective AI governance must recognize how consent challenges both influence and are influenced by considerations of fairness, transparency, accountability, and autonomy in an iterative relationship. This suggests that policy interventions must simultaneously address multiple principles rather than focusing narrowly on consent mechanics alone.

Generally, emerging regulatory approaches to consent in AI vary globally but tend to share principles of transparency, user control, and accountability. The EU's AI Act and GDPR emphasize explicit, informed consent for AI processing personal data, related legislation should be informed by this. The U.S. AI Bill of Rights advocates for user consent and opt-out mechanisms, while state-level laws like California's Consumer Privacy Rights Act enhance AI-related data protections. China's AI regulations require companies to obtain user consent before using personal data in algorithmic decision-making and India's Digital Personal Data Protection Act introduces similar consent-based protections. These regulations reflect a global push for ethical AI use while balancing innovation and individual rights.

However, recently we have been seeing a shift in AI regulation toward "simplification" and "flexibility" to foster "innovation and competitiveness". The EU is considering reviewing regulations and enforcement, including of the GPDR in the coming months, with these goals. In the U.S., policymakers are leaning toward self-regulation and sector-specific guidelines rather

---

[27] Facebook appealed the decision before the Higher Regional Court of Düsseldorf, which then referred the case to the Court of Justice of the European Union to clarify whether national competition authorities have the power to assess compliance with the GDPR and how certain provisions of the regulation should be interpreted and applied. (See Judgment of the Court (Grand Chamber) of 4 July 2023, C-252/21, available at:
https://curia.europa.eu/juris/document/document.jsf;jsessionid=896E5359CAB2BFA9C7DC178D9D901732?text=&docid=276478&pageIndex=0&doclang=en&mode=req&dir=&occ=first&part=1&cid=289301 ) than broad federal mandates. While fostering innovation and competitiveness may be a good thing, protection of fundamental rights including data privacy might be compromised.[28]

## What's an example of consent in AI done well? What's an example of consent in AI gone wrong?

Throughout the text we've seen examples of consent that have gone wrong, either because there has been no gathering of consent (e.g., Clearview AI), or because the multiple challenges that are present regarding data usage in AI got in the way (for example, the Target pregnant teen profiling).

Apple's approach to user data in its health and fitness features is a strong example of consent in AI done right. With HealthKit[29] and ResearchKit[30], Apple requires explicit user consent before collecting or sharing health data with third-party apps or researchers. Users can control what data is shared, with whom, and for what purpose, all within a transparent interface. This model prioritizes privacy while still enabling AI-driven insights, such as personalized health recommendations, without compromising user autonomy.

## What is missing to make consent work in AI?

Consent in AI is currently insufficient to protect people's data as it places too much responsibility on individuals while failing to account for the complexities of data collection, processing, and algorithmic decision-making. The current model relies on users reading and agreeing to lengthy, often opaque terms of service, which creates an illusion of choice rather than meaningful control. For consent to be effective, responsibility must extend beyond individuals to corporations, developers, and policymakers. Users alone cannot realistically enforce their rights, especially when AI systems operate in ways that are difficult to understand or challenge. The burden must shift towards those who create and deploy AI technologies, ensuring they proactively embed privacy, transparency, and fairness into their systems.

A significant shift in corporate and developer responsibility is needed to make AI consent functional. Organizations should design systems that minimize the need for constant user consent and instead adopt privacy-preserving defaults. Developers must prioritize explainability and accountability in their models, ensuring users understand how their data is used. Governments and regulatory bodies should enforce stricter requirements, making it mandatory for AI companies to demonstrate compliance with ethical and legal standards rather than assuming users will navigate these complexities on their own.

---

[28] A good reference on the importance of protecting privacy, how pervasive the collection of our data is, and how limited the protections we may have with regards to our behavior and the law is Carissa Véliz's Privacy is Power (2020).
[29] https://developer.apple.com/documentation/healthkit
[30] https://developer.apple.com/design/human-interface-guidelines/researchkit

To truly address these challenges, we need to consider how to reconcile the economic realities, power imbalances, and information asymmetries that shape AI-driven data processing. Policies must find ways to balance the rapid pace of AI innovation with the protection of fundamental rights, potentially requiring new regulatory models that go beyond existing frameworks. A more adaptive and user-friendly consent system must be developed—one that remains enforceable even as technology continues to evolve. Rethinking AI consent means not only empowering individuals but also ensuring corporate accountability, creating an environment where privacy and technological progress can coexist and reinforce one another.

## What needs to change?

Throughout this chapter, we have examined how the conventional notion of consent, while fundamental to data protection and privacy rights, proves increasingly insufficient in addressing the implications of AI-generated content derived from personal data. The historical evolution of consent (from medical ethics to digital privacy) has not kept pace with the transformative capabilities of generative AI technologies.

We have identified three fundamental challenges that strain traditional consent frameworks in AI contexts: the scope problem, the temporality problem, and the autonomy trap. These challenges are practical difficulties in implementation but also represent fundamental mismatches between conventional consent models and the technological realities of contemporary AI systems.

Moreover, we have shown how these consent limitations intersect with and amplify other core principles of responsible AI development and deployment, including fairness, transparency, accountability, and autonomy. The examples we've explored throughout this chapter demonstrate how AI-generated content creates new forms of personal representation that extend far beyond the scope of original consent, potentially undermining individual identity rights and autonomy in new ways.

As Flick (2009) argues, autonomy itself may not be a useful underlying principle on which to justify informed consent in information technology. Following Manson and O'Neill's (2007) approach, Flick proposes viewing informed consent as a "waiver of normative expectations" rather than an exercise of individual autonomy. This perspective shift could be particularly valuable for AI systems, where the traditional autonomy-centric model fails to address the practical realities of consent at scale. Such an approach would focus less on individual choice in each instance, and more on establishing clear normative boundaries that AI systems and their developers must respect.

Therefore, we believe that the notion of consent needs to evolve in both ethical and legal dimensions to reflect the technological reality we now inhabit. This evolution requires us to reconsider important questions about the relationship between individuals and their data, between autonomy and technological capability, and between legal protection and ethical responsibility.

Hoping to inspire further discussions within the interdisciplinary AI research community, we continue to wonder:

1. How might we reconceptualize consent in ways that acknowledge its inherent limitations in generative AI contexts while preserving its function as an expression of individual autonomy?
2. What alternative ethical frameworks might complement or partially replace consent when addressing AI systems whose outputs cannot be fully anticipated at the time of data collection?
3. How can we balance protecting individual rights with the collective benefits that may arise from AI systems trained on broad datasets?
4. What role should different stakeholders (from individual data subjects to AI developers, deployers, and regulators) play in addressing the consent gap?
5. How might we develop more dynamic models of consent that evolve alongside AI systems rather than representing static, one-time decisions?
6. What technical solutions might help bridge the gap between the limitations of traditional consent and the need for meaningful individual control over personal data and its derivatives?

The challenges we have outlined in this chapter are not only academic concerns, but have real-world implications for individual rights, social responsibility, and the future development of AI technologies. By examining the limitations of consent in AI contexts and raising these questions, we hope to contribute to an ongoing dialogue that leads to more ethically robust and legally sound approaches to AI governance.

# References

Bergemann, B. (2018). The Consent Paradox: Accounting for the Prominent Role of Consent in Data Protection. In: Hansen, M., Kosta, E., Nai-Fovino, I., Fischer-Hübner, S. (eds) Privacy and Identity Management. The Smart Revolution. Privacy and Identity 2017.

Bogucka, E., Constantinides, M., Šćepanović, S., & Quercia, D. (2024). AI Design: A Responsible AI Framework for Impact Assessment Reports. IEEE Internet Computing. DOI: http://doi.org/10.1109/MIC.2024.3451351

Bourtoule, Lucas, et al. "Machine unlearning." *2021 IEEE symposium on security and privacy (SP)*. IEEE, 2021.

Charlotte A. Tschider, Meaningful Choice: A History of Consent and Alternatives to the Consent Myth, 22. N.C. J.L. & TECH. 617 (2021).

Code of Nuremberg. "The Nuremberg code." *Trials of war criminals before the Nuremberg military tribunals under control council law* 10.1949 (1949): 181-2.

Desai, Deven R., and Mark Riedl. "Between Copyright and Computer Science: The Law and Ethics of Generative AI." Northwestern Journal of Technology and Intellectual Property 22, no. 1 (November 2024): 55-108.

Flick, Catherine. *Informed Consent in Information Technology: Improving User Experience*. PhD diss., Charles Sturt University, Centre for Applied Philosophy and Public Ethics, 2009.

Hasday, Lisa R. "The Hippocratic oath as literary text: a dialogue between law and medicine." Yale J. Health Pol'y L. & Ethics 2 (2001): 299.

IFIP Advances in Information and Communication Technology, vol 526. Springer, Cham. https://doi.org/10.1007/978-3-319-92925-5_8

Johnson, Deborah G. "Computer systems: Moral entities but not moral agents." *Ethics and information technology* 8 (2006): 195-204.

Longpre, Shayne, Robert Mahari, Ariel Lee, Campbell Lund, Hamidah Oderinwale, William Brannon, Nayan Saxena, et al. 2024. "Consent in Crisis: The Rapid Decline of the AI Data Commons" in *Proceedings of the 38th Conference on Neural Information Processing Systems (NeurIPS 2024) Track on Datasets and Benchmarks*.

Custers, B. H. M.. "Click here to consent forever: Expiry dates for informed consent." Big Data & Society 3 (2016).

Madianou, Mirca. *Technocolonialism - When Technology for Good is Harmful,* Polity Press (2025)

Manson, Neil C., and Onora O'Neill. *Rethinking Informed Consent in Bioethics*. Cambridge: Cambridge University Press, 2007.

Miller, F., & Wertheimer, A. (Eds.). (2009). *The ethics of consent: theory and practice*. Oxford University Press.

Muller, M., & Strohmayer, A. (2022, April). Forgetting practices in the data sciences. In Proceedings of the 2022 CHI Conference on Human Factors in Computing Systems (pp. 1-19). https://doi.org/10.1145/3491102.3517644

Muller, M. (2024). Data Silences: How to Unsilence the Uncertainties in Data Science. In Proceedings of the 2024 SIAM International Conference on Data Mining (SDM) (pp. 388-391). Society for Industrial and Applied Mathematics. DOI: https://doi.org/10.1137/1.9781611978032.44

Nozick, R. (1973). Distributive justice. *Philosophy & Public Affairs*, 45-126.

Rawls, J., *A Theory of Justice*, Cambridge, MA: Harvard University Press.

Schneble CO, Elger BS, Shaw DM. All Our Data Will Be Health Data One Day: The Need for Universal Data Protection and Comprehensive Consent. J Med Internet Res. 2020 May 28;22(5):e16879. doi: 10.2196/16879. PMID: 32463372; PMCID: PMC7290498.

Solove, Daniel J. Privacy Self-Management and the Consent Dilemma, 126 Harv. L. Rev. 1880 (2013).

Solove, Daniel J., Artificial Intelligence and Privacy (February 1, 2024). 77 Florida Law Review 1 (2025), GWU Legal Studies Research Paper No. 2024-36, GWU Law School Public Law Research Paper No. 2024-36, Available at SSRN: https://ssrn.com/abstract=4713111 or http://dx.doi.org/10.2139/ssrn.4713111

Solove, Daniel J. and Hartzog, Woodrow, The Great Scrape: The Clash Between Scraping and Privacy (July 03, 2024). 113 California Law Review (forthcoming 2025), Available at SSRN: https://ssrn.com/abstract=4884485 or http://dx.doi.org/10.2139/ssrn.4884485

Stahl, B. C., Antoniou, J., Bhalla, N., Brooks, L., Jansen, P., Lindqvist, B., ... & Wright, D. (2023). A systematic review of artificial intelligence impact assessments. Artificial Intelligence Review, 56(11), 12799-12831. DOI: https://doi.org/10.1007/s10462-023-10420-8

Tschider, Charlotte, *Meaningful Choice: A History of Consent and Alternatives to the Consent Myth*, 22 N.C. J.L. & Tech. 617 (2021).

United Nations General Assembly. The Universal Declaration of Human Rights (UDHR). New York: United Nations General Assembly, 1948

Warren, Samuel D., and Louis D. Brandeis. "The Right to Privacy." *Harvard Law Review*, vol. 4, no. 5, 1890, pp. 193–220. *JSTOR*, https://doi.org/10.2307/1321160. Accessed 24 Mar. 2025.

Xia, B., Lu, Q., Perera, H., Zhu, L., Xing, Z., Liu, Y., & Whittle, J. (2023, May). Towards concrete and connected AI risk assessment (C 2 AIRA): A systematic mapping study. In 2023 IEEE/ACM 2nd International Conference on AI Engineering–Software Engineering for AI (CAIN) (pp. 104-116). IEEE. DOI: http://doi.org/10.1109/CAIN58948.2023.00027

# Table of acronyms

ABOUT ML: Annotation and Benchmarking on Understanding and Transparency of Machine Learning Lifecycles

AI: Artificial Intelligence

AIA: Algorithmic Impact Assessment

EU: European Union

FIPPs: Fair Information Practice Principles

GDPR: General Data Protection Regulation

SAG-AFTRA: Screen Actors Guild - American Federation of Television and Radio Artist

SISA: Sharded, Isolated, Sliced, and Aggregated

UDHR: Universal Declaration of Human Rights

U.S.: United States of America